\documentclass[aps,prl,twocolumn,superscriptaddress,longbibliography]{revtex4-1}

\usepackage{amsmath}
\usepackage{graphicx}
\usepackage{bm}
\usepackage{amssymb}
\usepackage{hyperref}
\usepackage{xcolor}
\usepackage{amsmath}
\usepackage{amssymb}

\DeclareGraphicsExtensions{.png,.jpg,.eps}

\begin{document}
\author{Wei Luo}
\affiliation{National Laboratory of Solid State Microstructures, School of Physics,
and Collaborative Innovation Center of Advanced Microstructures, Nanjing University, Nanjing 210093, China}
\affiliation{School of Science, Jiangxi University of Science and Technology, Ganzhou 341000, China}

\author{Hao Geng}
\affiliation{National Laboratory of Solid State Microstructures, School of Physics,
and Collaborative Innovation Center of Advanced Microstructures, Nanjing University, Nanjing 210093, China}

\author{D. Y. Xing}
\affiliation{National Laboratory of Solid State Microstructures, School of Physics,
and Collaborative Innovation Center of Advanced Microstructures, Nanjing University, Nanjing 210093, China}

\author{G. Blatter}
\affiliation{Institute for Theoretical Physics, ETH Zurich, 8093 Zurich, Switzerland}

\author{Wei Chen}
\email{Corresponding author: pchenweis@gmail.com}
\affiliation{National Laboratory of Solid State Microstructures, School of Physics,
and Collaborative Innovation Center of Advanced Microstructures, Nanjing University, Nanjing 210093, China}
\affiliation{Institute for Theoretical Physics, ETH Zurich, 8093 Zurich, Switzerland}

\title{Entanglement of Nambu Spinors and Bell Inequality Test Without Beam
Splitters}

\begin{abstract} The identification of electronic entanglement in
solids remains elusive so far, which is owed to the difficulty of implementing
spinor-selective beam splitters with tunable polarization direction. Here, we
propose to overcome this obstacle by producing and detecting a particular type
of entanglement encoded in the Nambu spinor or electron-hole components of
quasiparticles excited in quantum Hall edge states. Due to the opposite charge
of electrons and holes, the detection of the Nambu spinor translates
into a charge-current measurement, which eliminates the need for beam
splitters and assures a high detection rate. Conveniently, the spinor
correlation function at fixed effective polarizations derives from a single
current-noise measurement, with the polarization directions of the detector
easily adjusted by coupling the edge states to a voltage gate and a
superconductor, both having been realized in experiments. We show that the
violation of Bell inequality occurs in a large parameter region. Our
work opens a new route for probing quasiparticle entanglement in solid-state
physics exempt from traditional beam splitters.  \end{abstract}

\date{\today}

\maketitle

%\emph{Introduction.}-
Appearing as a mystery in the early days of quantum mechanics~\cite{Einstein,schrodinger35},
quantum entanglement has %been demonstrated
%to be reality in modern physics \cite{Amico08rmp}.
%Moreover, it has
become a central resource of modern
quantum information science~\cite{bennett00nat,Nielsen},
which results in important applications in quantum computation, communication
and other protocols~\cite{Ekert,Bennett}. Despite its ubiquity in quantum
many-body systems, producing and manipulating entanglement in a controllable
way requires great efforts.
%that lies at the heart of quantum information science.
In practice, entangled states can be carried by different particles and various
physical degrees of freedom, which has stimulated diverse explorations in
different branches of physics. In particular, the manipulation of entangled
photons has turned into a mature technology~\cite{pan12rmp}; by contrast, the
detection of entanglement between quasiparticles in solids remains a challenge.

As an electronic analog of quantum optics, various
quantum interference effects have been implemented in mesoscopic
transport experiments~\cite{henny99sci,oliver99sci,ji03nat,neder07nat,weisz14sci}.
With this inspiration, numerous theoretical proposals have been made to
generate and detect spin and orbital entanglement
between separated electrons, or more precisely, quasiparticles,
in solid-state physics~\cite{lesovik01epjb,Recher01prb,Kawabata,
Chtchelkatchev,Samuelsson,Beenakker1,PhysRevLett.92.026805,beenakker05,Lebedev05prb,Samuelsson05prb,ChenW,chen13prb,Schroer,
Chirolli,Lorenzo,braunecker13prl}. Experimental progress has also been made
in the entanglement production via Cooper pair splitting in
hybrid superconducting structures~\cite{Hofstetter,Herrmann,wei10np,Hofstetter11prl,Schindele12prl,Das,Fulop,Tan15prl,Scher}.
%and electron-hole pair utilizing ac driving~\cite{Feve,Vanevi,Bisognin}.
Nevertheless, the verification of the expected entanglement
remains out of reach and
further development in this direction has been rather scarce in the past decade.
The main obstacle hindering the detection of spin/orbital entanglement
is the demand for high-quality spinor-selective beam splitters~\cite{Kawabata,
Chtchelkatchev,Samuelsson,Beenakker1,PhysRevLett.92.026805,beenakker05,Lebedev05prb,Samuelsson05prb,ChenW,chen13prb,Schroer,
Chirolli}.
Analogous to the polarisers used in optics,
which shall split particles with
orthogonal spinor states into distinct channels,
they play an essential role
in the entanglement detection such as
the two-channel Bell inequality (BI) test~\cite{Aspect2,Tittel}.
Moreover, the polarization of the beam splitters
should be adjustable, which is beyond
state-of-the-art techniques of solid-state physics.

\begin{figure}[t!]
\centering
\includegraphics[width=1\columnwidth]{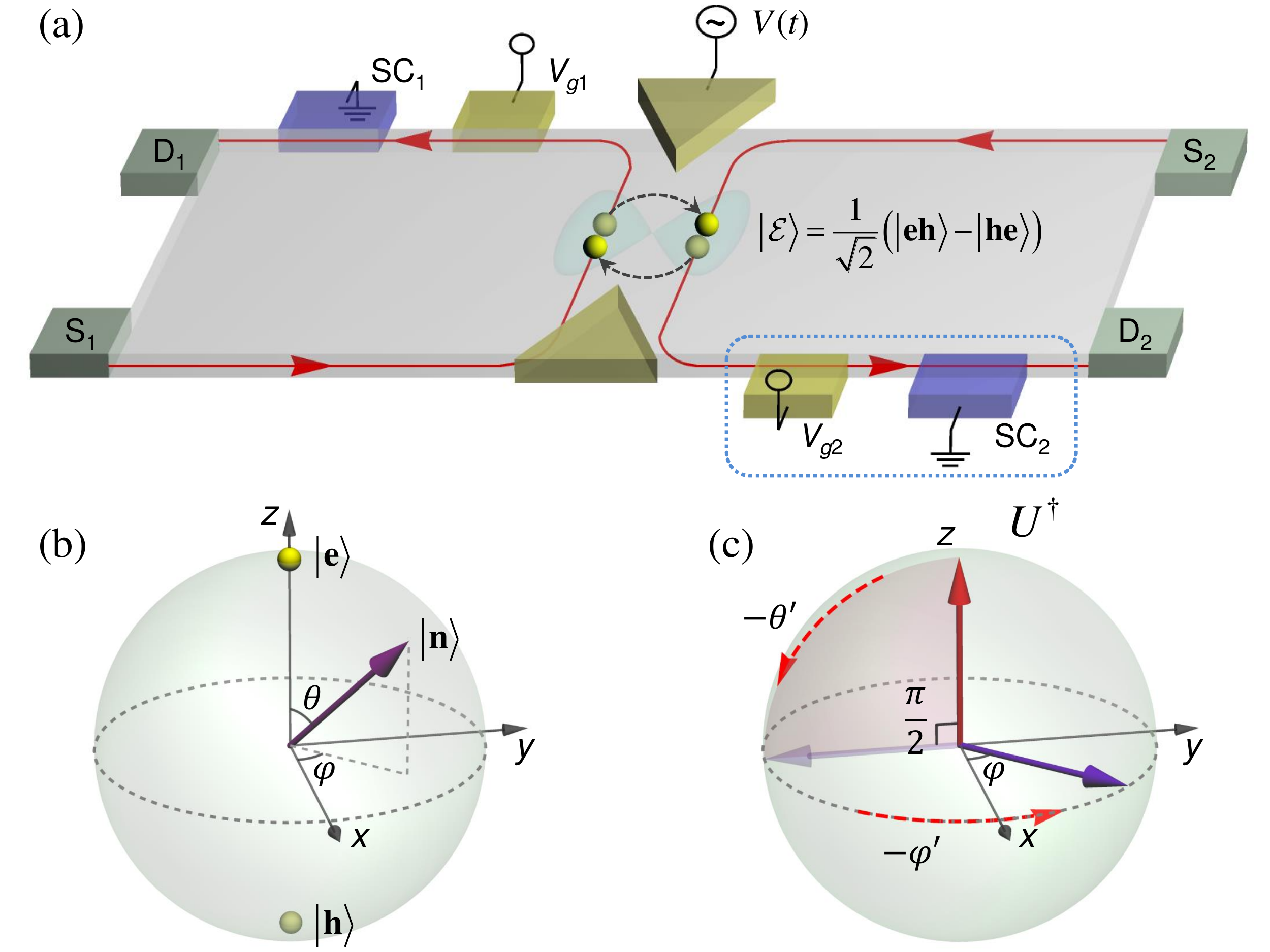}
\caption{(a) Proposed setup constructed
on the single chiral edge state (arrowed lines) in the integer quantum Hall regime.
Quasiparticle entanglement is generated
in the central region (hourglass area) by a
periodic electrical potential $V(t)$.
Two sources (S$_{1,2}$) and two drains (D$_{1,2}$) are all grounded.
A grounded superconductor (SC$_{1,2}$) and a
voltage gate ($V_{g1,2}$) are coupled to the outgoing
channel on each side.
(b) Representation of the Nambu spinor on the
Bloch sphere. The north and south poles correspond to the
electron ($|\textbf{e}\rangle$) and hole ($|\textbf{h}\rangle$) states, respectively.
(c) The actions of SC$_{1,2}$ and $V_{g1,2}$
inside the dashed box in (a) induce an
effective rotation
of the Nambu spinor.}
\label{Fig1}
\end{figure}

In this Letter, we propose to generate and detect a novel type of
entanglement encoded in the Nambu spinor (or the electron-hole qubits)
carried by quasiparticles excited in quantum Hall edge states,
see Fig.\ \ref{Fig1}(a). Taking advantage
of the \emph{opposite} charge carried by electrons
and holes, the spinor states can be detected
through a pure \emph{charge} measurement,
which successfully bypasses the need for
spinor-selective beam splitters.
We show that the BI test
with entangled Nambu spinors
can be implemented by measuring the charge current correlation, with
the polarization directions of the effective detectors on both sides
being adjusted by a voltage gate and a
superconductor coupled to the edge states, see Fig.\ \ref{Fig1}.
Given that the main ingredients of our proposal have all been realized
experimentally, the observation of a considerable violation of the BI
can be expected to come into reach. %Our work provides an exceptional example of
%solid-state superiority in certain
%quantum information processing, which stems from
%the unique property of the many-body Fermionic ground state.

%\emph{Entanglement of Nambu spinors.}-
The existence of the Fermi sea
at zero temperature allows one to
redefine it as the vacuum $|0\rangle$
of electron- and hole-like excitations.
The Nambu spinor is defined by the superposition
\begin{equation}\label{n}
|\textbf{n}\rangle=\cos\frac{\theta}{2}|\textbf{e}\rangle+e^{i\varphi}\sin\frac{\theta}{2}|\textbf{h}\rangle,
\end{equation}
where $|\textbf{e}(\textbf{h})\rangle=\gamma_{\text{e}(\text{h})}^\dag|0\rangle$
is the electron (hole) state created by the quasiparticle operator $\gamma^\dag_{\text{e}(\text{h})}$.
It can be visualised by the unit vector $\textbf{n}=(\sin\theta\cos\varphi,\sin\theta\sin\varphi,\cos\theta)$
on the Bloch sphere as shown in Fig.\ \ref{Fig1}(b).
In particular, the north and south poles represent the classical states $|\textbf{e}\rangle\!\Leftrightarrow\!|\uparrow\rangle$
and $|\textbf{h}\rangle\!\Leftrightarrow\!|\downarrow\rangle$,
which resemble the conventional spin-up and spin-down states, respectively.
The advantageous feature of the Nambu spinor $|\textbf{n}\rangle$
is that the electron and hole components have opposite charge, so that
the expectation value of the pseudospin $\tau_z$ (with $\tau_{x,y,z}$ the
Pauli matrices in Nambu space)
is given by $\langle\tau_z\rangle=\cos\theta=\langle Q\rangle/e$,
the average charge $\langle Q\rangle$ of the quasiparticle divided by
the elementary charge $e$. As a result,
the detection of the Nambu spinor along
the $z$-direction is equivalent
to measuring the charge that can be implemented \emph{locally}
without a beam splitter,
in stark contrast to the cases of the
spin and orbital states \cite{lesovik01epjb,Recher01prb,Kawabata,
Chtchelkatchev,Samuelsson,Beenakker1,PhysRevLett.92.026805,Lebedev05prb,Lorenzo,beenakker05,Samuelsson05prb,ChenW,chen13prb,braunecker13prl,Schroer,
Chirolli}.
Moreover, a perfect detection rate is also assured
thanks to the chiral edge states being immune to backscattering.

Next, we show that probing the Nambu spinor
along a general direction $\textbf{n}$, a precondition
for the entanglement detection, can
be achieved by coupling the edge state to
a voltage gate and a grounded superconductor [Fig.\ \ref{Fig1}(a)].
These physical elements induce two effective rotations of the Nambu spinor [Fig.\ \ref{Fig1}(c)],
with (i) the gate voltage $V_g$ generating a rotation
$U_G(\varphi')=\text{exp}\{-i\varphi'\tau_z/2\}$
about the $z$-axis.
This is owed to the opposite phase accumulated
by the electron and hole with magnitude $\varphi'=-2eV_gL_g/(\hbar v)$ given by
the length $L_g$ of the gated region and the velocity $v$ of the edge state.
And (ii), the superconductor causes Andreev reflection~\cite{Ostaay11prb,Beenakker14prl}
between electrons and holes
and introduces the other rotation
$U_{S}(\theta')=\text{exp}\{-i\theta'\tau_{x}/2\}$
about the $x$-axis by an angle $\theta'=2 \Delta L_s/(\hbar v)$
determined by the length $L_s$ and the effective pair potential $\Delta$
induced by the superconductor~\cite{SM}.
Very recently, such coherent electron-hole
conversion in the chiral edge states has been implemented in several experiments~\cite{Rickhaus12nl,LeeGH,ZhaoL,Hatefipour,Wangda}.
The detection of the Nambu spinor along the polarization direction $\textbf{n}$
means that the initial states $|\pm\textbf{n}\rangle$ should
yield the expectation values of $\pm1$ or explicitly,
$\langle\pm\textbf{n}|U^\dag\tau_z U|\pm\textbf{n}\rangle=\pm1$
with $U=U_S(\theta')U_G(\varphi')$ the combined rotational operation.
Given the direction of $\textbf{n}$, this requires rotation angles
$\theta'=-\theta$ and $\varphi'=-\varphi-\pi/2$,
which can be implemented by proper tuning of $V_g$ and $\Delta$
as shown in Fig.\ \ref{Fig1}(c).

%For systems with conserved electron number, the single-qubit state
%in Eq.\ \eqref{n} cannot be generated. However, the bipartite entangled
%state
%\begin{equation}\label{e}
%|\mathcal{E}\rangle=\frac{1}{\sqrt{2}}(|\textbf{e}\textbf{h}\rangle-|\textbf{h}\textbf{e}\rangle)
%\end{equation}
%can be implemented by noticing that creating a hole is equivalent to
%annihilating an electron of opposite energy. Thus, both terms in
%$|\mathcal{E}\rangle$ preserve the electron number.
%Such an entangled state resembles the spin-singlet state as $|\mathcal{E}\rangle\Leftrightarrow(|\uparrow\downarrow\rangle-|\downarrow\uparrow\rangle)/\sqrt{2}$.
The bipartite entanglement of the Nambu spinor
takes the form of
\begin{equation}\label{e}
|\mathcal{E}\rangle=\frac{1}{\sqrt{2}}(|\textbf{e}\textbf{h}\rangle-|\textbf{h}\textbf{e}\rangle),
\end{equation}
which resembles the spin-singlet state $|\mathcal{E}\rangle\!\Leftrightarrow\!
(|\!\uparrow\downarrow\rangle\!-\!|\!\downarrow\uparrow\rangle)\\/\sqrt{2}$.
We remark that the entanglement of the Nambu spinor~\cite{Strubi11prl,Chirolli}
is entirely \emph{different} from the previously proposed
electron-hole entanglement in Refs.~\cite{Beenakker1,PhysRevLett.92.026805,beenakker05}.
In the latter, the electrons and holes are merely physical carriers
of quantum information
while the qubit is still encoded in the spin or orbital degrees of freedom.
In contrast, here, the electron and hole components themselves constitute the qubit.

%\emph{Entanglement production.}-
The entangled state in Eq.\ \eqref{e}
can be prepared by the setup fabricated on
the quantum Hall edge states as shown in Fig.\ \ref{Fig1}(a).
The main ingredients include:
a central point-contact structure driven by
a periodic potential
$V(t)=V_{0}+V_{1}\cos{\omega t}$ of frequency $\omega$,
the sources S$_{j}$ and drains D$_{j}$
connected to the incident and outgoing channels,
and a voltage gate $V_{gj}$ and a superconductor SC$_{j}$ coupled to
the outgoing channels on each side ($j=1,2$). The many-body state of the electrons incident from
S$_1$ and S$_2$ reads
$|\Psi_{\mathrm{in}}\rangle=\prod_{\epsilon<0}\gamma_{1\text{e}}^{\text{in}\dagger}(\epsilon)\gamma_{2\text{e}}^{\text{in}\dagger}(\epsilon)|\rangle$,
with $\gamma_{1\text{e},2\text{e}}^{\text{in}\dagger}(\epsilon)$ the
creation operators of the incident electrons with an
energy $\epsilon$ measured from the Fermi level
and $|\rangle$ being the true vacuum of the electron.

Electron-hole pairs are excited around the point-contact region
by absorbing energy quanta $n\hbar\omega$ from the
periodic potential~\cite{feve2007demand,Gabor2009Science,Vanevi,Dubois2013Nat,bisognin2019quantum}.
%The physical process can be well described by the Floquet scattering
%matrix~\cite{Moskalets02prb,moskalets11} which relates the incident ($\gamma^{\text{in}}_{1\text{e},2\text{e}}$) and outgoing waves
%($\gamma_{1\text{e},2\text{e}}$) through
The physical process can be well described by the Floquet scattering
matrix~\cite{Moskalets02prb,moskalets11}, which relates the
incident ($\gamma^{\text{in}}_{1\text{e},2\text{e}}$) and outgoing
electron waves ($\gamma_{1\text{e},2\text{e}}$) through
\begin{equation}\label{s}
\begin{split}
  \left(
  \begin{array}{cc}
  \gamma_{1\text{e}}(\epsilon_{n})\\
  \gamma_{2\text{e}}(\epsilon_{n})\\
  \end{array}
  \right)=s(\epsilon_n,\epsilon)
  \left(
  \begin{array}{cc}
  \gamma^{\text{in}}_{1\text{e}}(\epsilon)\\
  \gamma^{\text{in}}_{2\text{e}}(\epsilon)\\
  \end{array}
  \right),\\
 s(\epsilon_{n},\epsilon)=\left(
  \begin{array}{cc}
  r(\epsilon_{n},\epsilon) & t'(\epsilon_{n},\epsilon)\\
  t(\epsilon_{n},\epsilon) & r'(\epsilon_{n},\epsilon)\\
  \end{array}
  \right),
  \end{split}
\end{equation}
where $t, t'$ ($r, r'$) are the
transmission (reflection) amplitudes,
$\epsilon_{n}=\epsilon+n\hbar\omega$
and the two components of the
column vector correspond to electrons on different sides.
In the limit of a weak driving potential $V_1$,
it is sufficient to consider only the static and single-photon-assistant scattering
processes described by $s_0=[r, t'; t, r']$
and $s_\pm=s(\epsilon\pm \hbar \omega, \epsilon)=[r_\pm, t'_\pm; t_\pm, r'_\pm]$, respectively.
Furthermore, the potential is assumed to vary slowly such that $\omega\delta t\ll1$
with $\delta t$ the time interval that the electron spends in
the central point-contact region.
In this adiabatic approximation, $s_{\pm}$
can be expressed in terms of
$s_0$ as $s_{\pm }=V_{1}(\partial s_0/\partial V_{1})/2$~\cite{Moskalets02prb}.

%Inserting Eq.\ \eqref{s} into the incident state
%$|\Psi_{\mathrm{in}}\rangle$ yields the outgoing wave function
%to first order in $V_{1}$,
By applying the complex conjugation of the relation Eq.~\eqref{s}
to the incident state $|\Psi_{\mathrm{in}}\rangle$,
one can obtain the outgoing wave function to the first order in $V_1$,
$
|\Psi_{\mathrm{out}}\rangle=|0\rangle+|{\tilde{\mathcal{E}}}\rangle+|\phi\rangle,
$
with
\begin{equation}\label{en}
|{\tilde{\mathcal{E}}}\rangle=\!\!\int_{-\hbar\omega}^{0}  \!\!\!\!\!\! d\epsilon \Big[f_{12}
\gamma_{1\text{e}}^{\dagger}(\epsilon_{1})\gamma_{2\text{h}}^{\dagger}(-\epsilon)-f_{21}
\gamma_{1\text{h}}^{\dagger}(-\epsilon)\gamma_{2\text{e}}^{\dagger}(\epsilon_{1})\Big]|0\rangle,
\end{equation}
the nonlocally entangled state in the form of Eq.~\eqref{e} composed
of quasiparticles with different energies, and
$
|\phi\rangle=\sum_{{j=1,2}}\int_{-\hbar\omega}^{0}d\epsilon f_{{jj}}\gamma_{{j}\text{e}}^{\dagger}(\epsilon_{1})
\gamma_{{j}\text{h}}^{\dagger}(-\epsilon)|0\rangle
$
the local electron-hole pairs.
The coefficients are defined by
$f_{11}=r_{+}r^{\ast}+t'_{+}{t'}^{\ast}$,
$f_{12}=r_{+}t^{\ast}+t'_{+}{r'}^{\ast}$,
$f_{21}=t_{+}r^{\ast}+r'_{+}{t'}^{\ast}$,
and $f_{22}=t_{+}t^{\ast}+r'_{+}{r'}^{\ast}$.
The unitarity of $s_0$ assures that $|f_{12}|=|f_{21}|$
and $|f_{11}|=|f_{22}|$, which results in
a maximal entanglement in Eq.\ \eqref{en}.
In the absence of $V_1$,
the vacuum state in terms of the
outgoing waves satisfies $|0\rangle=\prod_{\epsilon<0}\gamma_{1\text{e}}^{\dagger}(\epsilon)
\gamma_{2\text{e}}^{\dagger}(\epsilon)|\rangle=e^{i\delta}|\Psi_{\text{in}}\rangle$
which differs from the incident state by an unimportant overall phase $\delta$.
In the derivation of the outgoing state $|\Psi_{\text{out}}\rangle$,
the particle-hole transformation
$\gamma_{j\text{h}}^\dag(\epsilon)=\gamma_{j\text{e}}(-\epsilon)$
has been applied to the electrons undergoing
photon-assistant scattering.
The transition of an electron from energy
$\epsilon\in(-\hbar\omega,0)$ below the Fermi level
to $\epsilon_1\in(0,\hbar\omega)$ above
is equivalent to the creation of an electron-hole pair
in the new vacuum $|0\rangle$.

%\emph{Shot noise and BI test.}-
The entanglement
of Nambu spinors can be
measured by the charge current correlator.
As discussed previously, the drain D$_{j}$
together with the gate voltage $V_{gj}$ and the superconductor SC$_{j}$
comprise an effective spinor detector
along the polarization direction $\textbf{n}_{j}$ [cf. Fig.\ \ref{Fig1}].
Mathematically, this effect can be absorbed into the modified current operator
in D$_j$ as
$\hat{I}_{\textbf{n}_{j}}(t)\!\!=\!\!\frac{e}{h}\int\!\!\int d\epsilon d\epsilon'e^{\frac{i}{\hbar}(\epsilon-\epsilon')t}
\gamma_{j}^\dag(\epsilon) U^\dag_{\textbf{n}_{j}}\tau_zU_{\textbf{n}_{j}}\gamma_{j}(\epsilon')$,
with $\gamma_j=(\gamma_{j\text{e}},\gamma_{j\text{h}})^{\text{T}}$
and the rotational operator $U_{\textbf{n}_j}=U_S(-\theta_j)U_G(-\varphi_j-\pi/2)$ specified by
the unit vector $\textbf{n}_j=(\sin\theta_j\cos\varphi_j,\sin\theta_j\sin\varphi_j,\cos\theta_j)$~\cite{SM}.
Without the actions of the gate voltage and the superconductor,
we have $\textbf{n}_j=\hat{z}$ and $U_{\textbf{n}_j}=\tau_0$
(unit operator) so that the electron and hole contribute oppositely to the charge current
which naturally measures the pseudospin $\tau_z$.
%It indicates that the conventional charge current operator,
%$\hat{I}_{\textbf{n}_j}(t)$ measures the spinor current
%in Nambu space at the same time, which is the key
%advantage of our scheme.

With the two sources S$_{1,2}$ being grounded, an
instantaneous current is generated by the
driving potential $V_1$.
Nevertheless, the average current associated with $|\Psi_{\text{out}}\rangle$
vanishes, i.e., $\langle\hat{I}_{\textbf{n}_j}\rangle$=0~\cite{footnote}.
Despite the zero average current,
the outgoing wave can give rise to finite current fluctuations.
Specifically, the zero-frequency noise power between the two drains D$_{1,2}$ is
defined by $\mathcal{S}(\textbf{n}_1,\textbf{n}_2)=2\int_{-\infty}^{\infty}
dt\langle\delta \hat{I}_{\textbf{n}_1}(t)\delta \hat{I}_{\textbf{n}_2}(0)\rangle$
with $\delta \hat{I}_{\textbf{n}_j}(t)=\hat{I}_{\textbf{n}_j}(t)-\langle \hat{I}_{\textbf{n}_j}\rangle$.
It turns out that only $|\tilde{\mathcal{E}}\rangle$
in $|\Psi_{\text{out}}\rangle$ contributes to the noise power,
while $|\phi\rangle$ contains neutral electron-hole
pairs on the same side that do not induce nonlocal correlation.
Therefore, the noise power reduces to $\mathcal{S}(\textbf{n}_1,\textbf{n}_2)=2\int_{-\infty}^{\infty}
dt\langle\tilde{\mathcal{E}}| \hat{I}_{\textbf{n}_1}(t) \hat{I}_{\textbf{n}_2}(0)|\tilde{\mathcal{E}}\rangle$
and provides a pure signature of the bipartite entanglement.
A straightforward calculation yields
\begin{equation}\label{noise}
\begin{split}
\mathcal{S}(\textbf{n}_1,\textbf{n}_2)
=\mathcal{S}_{\text{ee}}-\mathcal{S}_{\text{eh}}-\mathcal{S}_{\text{he}}+\mathcal{S}_{\text{hh}}
=\mathcal{S}_0\mathcal{P}(\textbf{n}_1,\textbf{n}_2),
\end{split}
\end{equation}
where the noise power is proportional to
the spinor correlation function $\mathcal{P}(\textbf{n}_1,\textbf{n}_2)=\langle\mathcal{E}
|(\bm{\tau}_1\cdot\textbf{n}_1)(\bm{\tau}_2\cdot\textbf{n}_2)|\mathcal{E}\rangle=-\textbf{n}_1\cdot\textbf{n}_2$
with the prefactor $\mathcal{S}_0=2e^{2}\omega|f_{12}|^{2}/\pi$ the sum of the four terms,
which satisfy $\mathcal{S}_{\text{ee}}=\mathcal{S}_{\text{hh}}$,
$\mathcal{S}_{\text{eh}}=\mathcal{S}_{\text{he}}$.
As a result, the spinor correlation function
can be measured directly by a \emph{single} noise power.
In the previous proposals based on spin and orbital
entanglements, the four spinor-resolved
noise powers $\mathcal{S}_{\tau\tau'}$ ($\tau,\tau'=\uparrow,\downarrow$)
needed to be measured separately and then
combined properly as in Eq.\ \eqref{noise}
to arrive at the spinor correlation $\mathcal{P}(\textbf{n}_1,\textbf{n}_2)$
~\cite{Kawabata,
Chtchelkatchev,Samuelsson,Beenakker1,PhysRevLett.92.026805,beenakker05,Lebedev05prb,Samuelsson05prb,ChenW,
Lorenzo,braunecker13prl}. This is out of reach in the experiment
not only because of its complexity in practical operations
but is also impeded by the necessity for high-quality spinor-selective beam
splitters. Remarkably, the four terms of the noise power naturally
appear with correct signs in Eq.\ \eqref{noise}, which greatly facilitate
the entanglement detection and improve the accuracy of measurement.

The violation of the BI involves a standard
criterion for the identification of quantum entanglement.
We adopt the Bell-Clauser-Horne-Shimony-Holt
form of the inequality~\cite{bell64,PhysRevLett.23.880}
\begin{equation}\label{BI}
\mathcal{B}=|\mathcal{P}(\textbf{n}_1,\textbf{n}_2)-\mathcal{P}(\textbf{n}_1,\tilde{\textbf{n}}_2)+\mathcal{P}(\tilde{\textbf{n}}_1,\textbf{n}_2)+\mathcal{P}(\tilde{\textbf{n}}_1,\tilde{\textbf{n}}_2)|\leq2,
\end{equation}
which contains four spinor correlation
functions along different
polarization directions. Note that
the correlation functions are measured by the noise
power $\mathcal{S}(\textbf{n}_1,\textbf{n}_2)$
via the relation in Eq.\ \eqref{noise}.
Specifically, the polar angle $\theta_j$
and the azimuthal angle $\varphi_j$ of the spinor ``polariser'' are controlled
by the coupling to the superconductor and the gate voltage through
$\theta_j=-2\Delta_j L_{s}/(\hbar v)$ and $\varphi_j=2eV_{gj}L_{g}/(\hbar v)-\pi/2$, respectively,
where we have assumed that the superconductors (gating regions)
on the two sides are of the same length, $L_s$ ($L_g$).
%The spinor correlation function is
%\begin{equation}
%\begin{split}
%\mathcal{P}(\textbf{n}_1,\textbf{n}_2)
%=-\cos{\theta_1}\cos\theta_2-\cos{(\varphi_{1}-\varphi_{2})}\sin{\theta_1}\sin\theta_2
%\end{split}
%\end{equation}

\begin{figure}[t!]
\centering
\includegraphics[width=1\columnwidth]{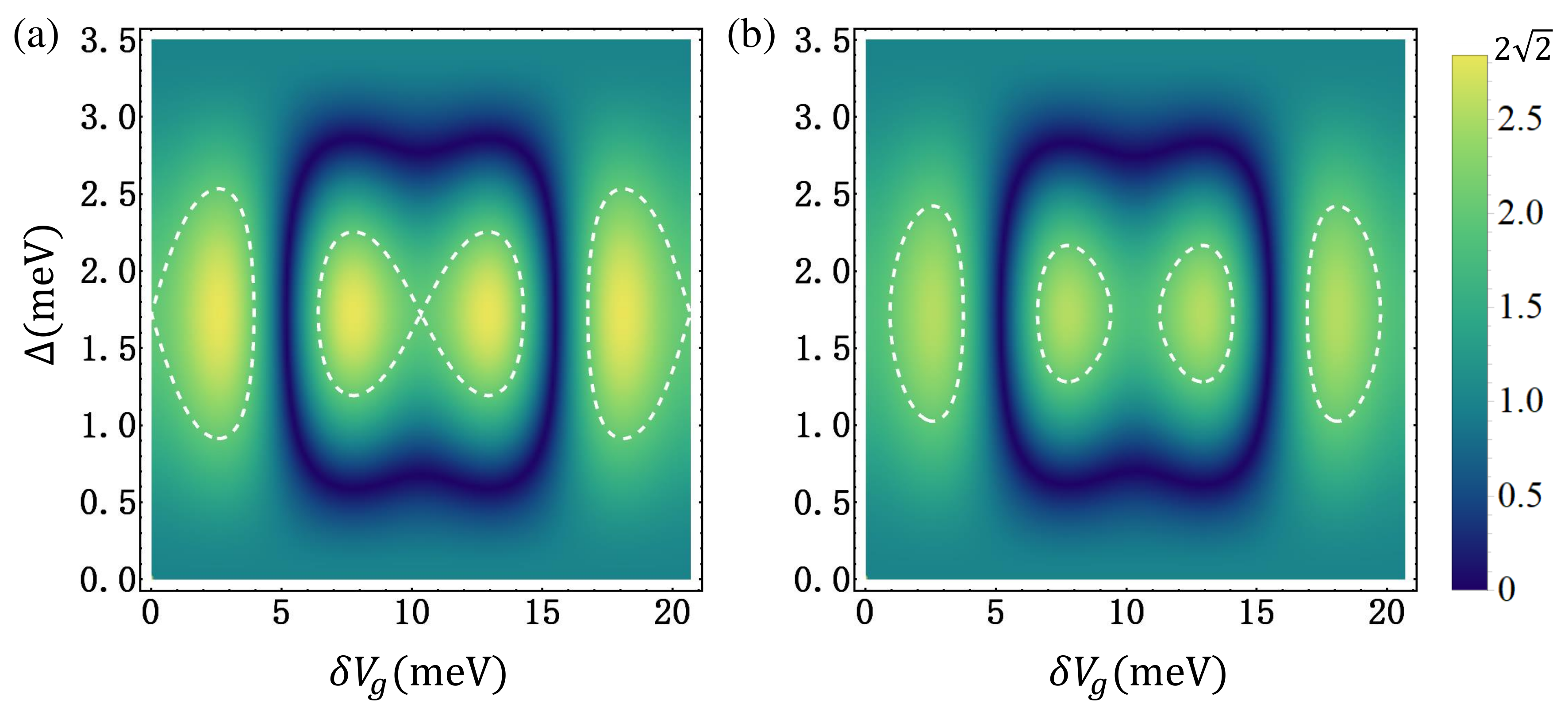}
\caption{Plot of $\mathcal{B}$ as a function of $\Delta$ and $\delta V_{g}$ with
(a) $\lambda=1$ and (b) $\lambda=0.9$. The dashed lines are the critical contours
of $\mathcal{B}=2$.
The relevant parameters are set as
$v=1\times10^{6}\mathrm{m/s}$, $L_s=300\mathrm{nm}$, and $L_g=100\mathrm{nm}$.}
\label{Fig2}
\end{figure}

For $\theta_1=\theta_2=\theta$,
the violation of the BI can be tested by taking the configuration of
the azimuthal angles as
$\varphi_{1}-\varphi_{2}=\tilde{\varphi}_{1}-\tilde{\varphi}_{2}=\varphi_{2}-\tilde{\varphi}_{1}=\delta\varphi=2e\delta V_{g}L_{g}/(\hbar v)$,
which can be regulated simply by the gate voltages $V_{g1,2}$ on both sides.
Assuming experimentally realizable parameters for the setup [see caption
of Fig.\ \ref{Fig2}], that can be achieved by
modern electron-beam lithography techniques~\cite{henny99sci,oliver99sci,ji03nat,neder07nat,weisz14sci,Rickhaus12nl,LeeGH,ZhaoL,Hatefipour,Wangda},
we plot $\mathcal{B}$ as a function of $\Delta$ and $\delta V_g$
in Fig.\ \ref{Fig2}(a).
One can see that the violation of the BI
occurs in a large parameter region encircled by the
critical contour $\mathcal{B}=2$ (dashed
lines).
The maximal violation $\mathcal{B}=2\sqrt{2}$ takes place when $\Delta=1.73$ meV,
and $V_{g1}-V_{g2}=\tilde{V}_{g1}-\tilde{V}_{g2}=V_{g2}-\tilde{V}_{g1}=\delta V_{g}=2.59$ meV
which corresponds to $\theta=\pi/2$ and $\delta\varphi=\pi/4$, respectively
(for results on more general cases with $\theta_1\neq\theta_2$
see Supplemental Material \cite{SM}.)

%\emph{Dephasing effect.}-
The detection of
entanglement requires the system
to preserve phase coherence. In reality, dephasing
dominated by energy averaging and temperature smearing
is present in the quantum Hall edge state~\cite{ji03nat,neder07nat}.
Then the entanglement
should be described by the density matrix
$\hat{\rho}=[|\textbf{e}\textbf{h}\rangle\langle \textbf{e}\textbf{h}|
+|\textbf{h}\textbf{e}\rangle\langle \textbf{h}\textbf{e}|
-\lambda(|\textbf{e}\textbf{h}\rangle\langle \textbf{h}\textbf{e}|
+|\textbf{h}\textbf{e}\rangle\langle \textbf{e}\textbf{h}|)]/2$
with $\lambda\in [0,1]$ the dephasing factor
which
can be estimated by $\lambda\simeq e^{-(L_g+L_s)/L_\phi}$
with $L_\phi$ the phase coherence length.
Accordingly, the crossed current correlation
is evaluated by $\langle\delta \hat{I}_{\textbf{n}_1}(t)\delta
\hat{I}_{\textbf{n}_2}(0)\rangle=\mathrm{Tr}[\hat{\rho} \delta\hat{I}_{\textbf{n}_1}(t)\delta\hat{I}_{\textbf{n}_2}(0)]$
and the spinor correlation function becomes
$\mathcal{P}'(\textbf{n}_1,\textbf{n}_2)=-\cos{\theta_1}\cos\theta_2
-\lambda\cos{(\varphi_{1}-\varphi_{2})}\sin{\theta_1}\sin\theta_2$.
The BI parameter $\mathcal{B}$ as a function of
$\Delta$ and $\delta V_g$ is plotted in Fig.\ \ref{Fig2}(b).
One can see that the areas of BI violation maintain, along with a certain shrinkage.
The BI can be violated as long as $\lambda>1/\sqrt{2}$, similar
to the situation of the two-qubit Werner states \cite{Werner89pra,Horodecki09rmp,Brunner14rmp}.

Sufficiently long coherence length and two-particle interference have both
been realized in mesoscopic devices constructed on quantum Hall edge
states \cite{ji03nat,neder07nat}.  A visibility as high as 90\% of the
oscillating pattern of the quantum Hall interferometer (with a path length
$\sim4$ $\mu$m) has been implemented \cite{ji03nat}, which indicates an
associated value $\lambda\simeq0.9$. Therefore, a considerable violation of
the BI can be achieved with the proposed length scales $L_s$ and $L_g$ given
in Fig. \ref{Fig2}. Moreover, the dephasing processes that occur after the
wave packet is transmitted through the voltage gate and superconducting region
does not affect the entanglement detection.  The reason is that the action of
the random phase modulation commutes with $\tau_z$ and thus cancels out its
complex conjugation in the current operator $\hat{I}_{\textbf{n}_{j}}(t)$.  As
a result, only the lengths $L_g$ and $L_s$ are of importance for coherent
transport, that further relaxes the constraint on the scale of the setup by
$L_\phi$.

%\emph{Parameter calibration and experimental implementation.}-
We specify the experimental procedures in detail.
The polarization direction $\textbf{n}_j$ of the spinor detector
can be adjusted by the gate voltage $V_{gj}$ and the Andreev reflection
at the superconductor SC$_j$~\cite{Rickhaus12nl,LeeGH,ZhaoL,Hatefipour,Wangda}.
It is convenient to place the four vectors $\textbf{n}_j, \tilde{\textbf{n}}_j$
in Eq.\ \eqref{BI} to the $x$-$y$ plane, i.e.,
$\theta_j=\pi/2$ [cf. Fig.\ \ref{Fig1}(c)] and change the azimuthal angles $\varphi_j$
by $V_{gj}$.
This corresponds to an Andreev reflection with 50\% probability, which can be
realized by tuning the magnetic field or a large
backgate~\cite{Rickhaus12nl,LeeGH,ZhaoL,Hatefipour,Wangda}.
To confirm the correct setting, one can exert
a large potential $V_0$ at the central point contact
to pinch off the connection between the edge channels on
both sides. Then by imposing a small bias $V_{sj}\ll\Delta_j$
to the source S$_j$ and measuring the
current $I_{dj}$ in the drain D$_j$,
one can verify $\theta_j=\pi/2$
by $I_{dj}=0$~\cite{SM}.

The quantitative relation between the azimuthal angle $\varphi_j$
and the gate voltage $V_{gj}$ should be established as well.
To implement this, four auxiliary point contacts G$_{1\text{-}4}$ are fabricated
to construct a Mach-Zehnder interferometer on each side \cite{ji03nat,neder07nat},
see Fig.\ S.2 in the Supplemental Material \cite{SM}. The whole setup is still
pinched off at the center by a large $V_0$. By fitting the oscillation pattern
modulated by $V_{gj}$
one can obtain the angle $\varphi_j$ as its function.
A linear dependence $\varphi_j\propto V_{gj}$ usually
holds in the regime of interest~\cite{ji03nat,neder07nat},
which facilitates the calibration of $\varphi_j$.

The prefactor $\mathcal{S}_0$ in Eq.\ \eqref{noise} should be probed so as to
normalize the spinor correlation function.  Note that without the
superconductor, the noise power equals
$\mathcal{S}(\hat{z},\hat{z})=-\mathcal{S}_0$.  Therefore, $\mathcal{S}_0$ can
be measured directly by the noise power between the two auxiliary drains
D$_{3,4}$ with the auxiliary point contacts G$_{2,4}$ being pinched
off~\cite{SM}.  In the measurement of $\mathcal{S}_0$ one first reduces $V_0$
to allow tunneling between the edge states on different sides and then imposes
the driving potential $V_1$.  The same setting of $V_0$ and $V_1$ should be
used during the Bell measurement.  With the calibration of all the
parameters discussed above, the spinor correlation can be determined
through $\mathcal{P}(\textbf{n}_1,\textbf{n}_2)
=\mathcal{S}(\textbf{n}_1,\textbf{n}_2)/\mathcal{S}_0$.  After completing
these calibrations, the gate voltages at the
auxiliary point contacts should be removed
to separate the edge channels on opposite sides
during the entanglement detection \cite{SM}, and the whole setup reduces to that in
Fig.~\ref{Fig1}.

One remaining point that needs clarification
is the existence of several undetermined phases. They are:
%(i) The initial phase factor $\gamma_1$ involved in the oscillating
%potential $V(t)=V_0+V_1\cos(\omega t+\gamma_1)$;
(i) The residual phases for the azimuthal angles $\varphi_j$ at
$V_{gj}=0$ accumulated during free propagation, which we denote
by their difference $\gamma_{1}=(\varphi_1-\varphi_2)|_{V_{gj}=0}$;
(ii) The $U(1)$ phases of the superconducting order parameters
of SC$_{1,2}$ whose difference
is denoted by $\gamma_2=\arg (\Delta_1)-\arg (\Delta_2)$;
(iii) The phase difference between the two coefficients
in Eq.\ \eqref{en} denoted by $\gamma_3=\arg (f_{12})-\arg (f_{21})$.
In our previous discussion,
all these phases have been chosen to be zero for simplicity.
We remark that these nonzero phases $\gamma_{1,2,3}$
do not affect the entanglement detection
and the violation of the BI.
Specifically, they only
introduce an overall phase shift
to the spinor correlation as
$\mathcal{P}(\textbf{n}_1,\textbf{n}_2)=-\cos{\theta_1}\cos\theta_2
-\cos{(\varphi_{1}-\varphi_{2}+\sum_{l=1}^3\gamma_l)}\sin{\theta_1}\sin\theta_2$,
which can be simply absorbed into $\varphi_1$ or $\varphi_2$.
For the entanglement detection,
the absolute values of $\varphi_{1,2}$ are not important,
but only their differences~\cite{foot}.

%\emph{Summary and outlook.-}
To summarize, we propose to detect the entanglement of quasiparticles by
encoding the qubit in the electron and hole
components of the Nambu spinor. The effective spinor correlator can be
measured directly by a single noise power function without spinor-selective
beam splitters and the BI can be considerably violated.  Our scheme
is pertaining to condensed matter physics, where the unique many-body
Fermionic ground state allows for the definition of an
electron-hole spinor, that has no optical counterpart.  The entangled Nambu
spinors can be used to explore other quantum correlation effects and may lead
to interesting applications in quantum information processing.

\begin{acknowledgments}
This work was supported by the National Natural Science Foundation of
China under Grant No. 12074172 (W.C.) and No. 11804130 (W.L.), the startup
grant at Nanjing University (W.C.), the State Key Program for Basic
Researches of China under Grants No. 2017YFA0303203 (D.Y.X.) the Excellent
Programme at Nanjing University and the financial support from the Swiss
National Science Foundation (SNSF) through Division II and QSIT (W.C. and G.B.).
\end{acknowledgments}

%\bibliography{Entanglement}

\begin{thebibliography}{59}
\bibitem [{foot()}]{foot}%
  \BibitemOpen
  \href@noop {} {}\bibinfo {note} {All these unknown phases
are determined by the specific conditions
of the setup; they are not subject to fluctuations
and thus will not destroy the phase coherence.}\BibitemShut {Stop}%
\end{thebibliography}
%\newpage
%\onecolumngrid
%\renewcommand{\theequation}{S.\arabic{equation}}
%\setcounter{equation}{0}
%\renewcommand{\thefigure}{S.\arabic{figure}}
%\setcounter{figure}{0}
%

\end{document}